\documentclass[12pt]{article}
\usepackage{a4wide}
\usepackage{amssymb}
\usepackage{xcolor}
\usepackage{hyperref}
\usepackage[normalem]{ulem}

\newcommand{\bea}{\begin{eqnarray}}
\newcommand{\eea}{\end{eqnarray}}

\def\[{\left[}
\def\]{\right]}
\def\({\left(}
\def\){\right)}
\def\nn{\nonumber}
\def\d{{\rm d}}
\def\A{\mathcal{A}}

\begin{document}
{\renewcommand{\thefootnote}{\fnsymbol{footnote}}
\begin{center}
{\LARGE  Signature change in loop quantum gravity:\\ General midisuperspace
  models and dilaton gravity}\\
\vspace{1.5em}
Martin Bojowald$^1$\footnote{e-mail address: {\tt
      bojowald@gravity.psu.edu}}
  and Suddhasattwa Brahma$^{1,2}$\footnote{e-mail address: {\tt suddhasattwa.brahma@gmail.com}}\\
  \vspace{0.5em}
$^1$  Institute for Gravitation and the Cosmos,\\
  The Pennsylvania State
  University,\\
  104 Davey Lab, University Park, PA 16802, USA\\
\vspace{0.5em}
$^2$ Center for Field Theory and Particle Physics,\\
Fudan University, 200433 Shanghai, China\\
\vspace{1.5em}
\end{center}
}

\setcounter{footnote}{0}

\begin{abstract}
  Models of loop quantum gravity based on real connections have a
  deformed notion of general covariance, which leads to the phenomenon
  of signature change. This result is confirmed here in a general
  analysis of all midisuperspace models without local degrees of
  freedom. As a subclass of models, 2-dimensional theories of dilaton
  gravity appear, but a larger set of examples is possible based only
  on the condition of anomaly freedom. While the classical dilaton
  gravity models are the only such systems without deformed
  covariance, they do give rise to signature change when holonomy
  modifications are included.
\end{abstract}

\section{Introduction}

In canonical formulations of gravitational theories, covariance is ensured by
gauge transformations generated by the constraints rather than by coordinate
transformations. Poisson brackets of the constraint functions on phase space
must then obey a certain form that reduces to the hypersurface deformations of
general relativity in the classical limit. Anomaly freedom, or the fact that
the constraints in modified or quantized gravity models must remain first
class, imposes strong conditions on the possible forms of constraints and on
structure functions in their brackets. Signature change is the most
characteristic and apparently generic consequence of these conditions.

Conditions that ensure a canonical quantum theory of gravity to be covariant
have been formulated in \cite{SphSymmCov}. It has been shown that not only (i)
the classical Hamiltonian and diffeomorphism constraints, on quantisation,
must still satisfy a first-class system and have a closed algebra; but also
(ii) that this algebra must have a classical limit whereby it reduces to the
familiar hypersurface deformation algebra \cite{DiracHamGR,ADMre} of general
relativity. This statement holds also for effective or modified theories in
which certain quantum corrections are included while working in a
semiclassical approximation. Covariance therefore poses an important
consistency question for canonical quantum gravity theories, which goes beyond
the requirement that constraints be anomaly free. Brackets (or commutators) of
the constraints not only have to lead to a closed system, they must also close
in such a way that a specific classical limit is obtained.

The examples discussed in detail in \cite{SphSymmCov} show that anomaly
freedom of gravitational models does not necessarily imply covariance. In
particular, constraint brackets in midi\-superpace models can often be
simplified by redefining the classical constrained system, sometimes
eliminating structure functions. The resulting Lie algebras are then easier to
quantize in an anomaly-free way. However, after quantization, it is not
guaranteed that the redefinitions can still be inverted such that a closed set
of hypersurface-deformation generators is obtained. The main example given in
\cite{SphSymmCov} is a partially Abelianized redefinition along the lines of
\cite{LoopSchwarz}, which can be made covariant in the presence of holonomy
modifications but only if there is no matter coupled to the system. Moreover,
signature change is realized also in the partially Abelianized system if
holonomy modifications are present.

Recently, several other models have been analyzed by partial
Abelianization, with proposed quantizations. In \cite{GowdyAbel}, a
locally rotationally symmetric Gowdy model has been introduced and
quantized in this way. In \cite{AbelDil}, the class of 2-dimensional
dilaton gravity models has been studied, with a special discussion of
the vacuum CGHS model \cite{CGHS} given in \cite{AbelCGHS}. These
models do not have local degrees of freedom and therefore do not
encounter the obstructions found in \cite{SphSymmCov,GowdyCov} for
covariant holonomy-modified models with local degrees of
freedom. Nevertheless, the question of covariance has not been
addressed in \cite{GowdyAbel,AbelDil,AbelCGHS}. We will fill in this
lacuna in the present paper. At the same time, we construct the most
general covariant midisuperspace model without local degrees of
freedom with spatial derivatives of the metric (or dyad and dilaton)
up to second order. We compute the modified structure functions of all
these models and conclude that the class of all classical
2-dimensional dilaton gravity models, with an arbitrary dilatonic
potential but the same form of the extrinsic curvature type components
as in general relativity, is the only set with undeformed
covariance. However, a large class of covariant models exists with
deformed covariance, which includes quantum versions of these dilaton
models with effects from loop quantum gravity. Most of these new
models have signature change if modification functions are such that
they mimic holonomy modifications of loop quantum gravity.

\section{Signature change in polarized Gowdy model with local rotational
  symmetry}

We first look at the specific model studied in \cite{GowdyAbel}: the polarised
Gowdy model on a three-torus with local rotational symmetry (LRS). The last
condition eliminates local degrees of freedom. As usual, we identify the two
homogeneous directions, $x$ and $y$, with each other while keeping the
inhomogeneous direction $\theta$ unchanged. We have an inhomogeneous
midisuperspace model without local physical degrees of freedom.

In keeping with the conventions of \cite{GowdyAbel}, we work with the two
triad components $\(E^x, \varepsilon\)$ and the extrinsic curvature components
conjugate to them, $\(K_x, \mathcal{A}\)$. In the reduced 1-dimensional
manifold with coordinate $\theta$, $E^x$ and ${\cal A}$ have density weight
one. The Poisson brackets between the canonical variables are
$\left\{K_x\(\theta_1\), E^x\(\theta_2\)\right\} =
G\delta(\theta_1,\theta_2) = \left\{\mathcal{A}\(\theta_1\),
  \varepsilon\(\theta_2\)\right\}$. Derivatives with respect to the
inhomogeneous coordinate are labelled by primes in the following.

As in the well-known case of spherical symmetry, there is only one global
degree of freedom. However, the form of the Hamiltonian constraint in the
Gowdy LRS case is distinct from that of spherical symmetry due to a different
internal curvature term.  For the latter model, the constraint is given by
\begin{eqnarray}\label{HamClass}
H[N] &=& -\frac{1}{2G}\int \d\theta N(\theta)\[\varepsilon^{-1/2}
K_x^2E^x + 4\varepsilon^{1/2}\mathcal{A}K_x
+\frac{1}{4}\varepsilon^{-1/2}\(E^x\)^{-1}\(\varepsilon^\prime\)^2\right.\nn\\
& &\hspace{15mm}
\left. +\varepsilon^{1/2}\varepsilon^{\prime\prime}\(E^x\)^{-1} -
  \(E^x\)^{-2}\varepsilon^{1/2}\varepsilon^{\prime}\(E^x\)^{\prime}\]\,,
\end{eqnarray}
while  the diffeomorphism constraint
\begin{eqnarray}\label{Diffeo}
D[N^x] &=& \frac{1}{G}\int \d\theta N^x(\theta) \[K_x^\prime E^x -
\varepsilon^\prime\mathcal{A}\]
\end{eqnarray}
takes the same form as in spherically symmetric models.  The classical
constraint brackets take the form
\begin{eqnarray}\label{ClassicalAlg}
\{D[N^\theta_1], D[N^\theta_1]\} &=& D\[\mathcal{L}_{N^\theta_1} N^\theta_2\]\\
\{H[N], D[N^\theta]\} &=& -H\[\mathcal{L}_N N^\theta\]\\
\{H[N_1], H[N_1]\} &=& D\[q^{\theta\theta}\(N_1N^\prime_2 - N_2N^\prime_1\)\]\,.
\end{eqnarray}
The only non-constant structure function
$q^{\theta\theta}=\varepsilon/(E^x)^2$ appears in the classical algebra above,
while the other non-zero components of the inverse spatial metric are $g^{xx}
= g^{yy} =\varepsilon^{-1}$. (It follows from the results of \cite{SphSymmPSM}
that the Hamiltonian constraint (\ref{HamClass}) is the same as what is
obtained for a 2-dimensional dilaton gravity model with zero dilaton
potential, when expressed in connection variables after a canonical
transformation. The LRS Gowdy model of \cite{GowdyAbel} is therefore nothing
but a CGHS model with zero cosmological constant.)

We introduce holonomy modifications in the Hamiltonian constraint
\begin{eqnarray}\label{Ham}
H[N] &=& -\frac{1}{2G}\int \d\theta N(\theta)\[\varepsilon^{-1/2}
f_1\(K_x\)E^x + 4\varepsilon^{1/2}\mathcal{A}f_2(K_x)
+\frac{1}{4}\varepsilon^{-1/2}\(E^x\)^{-1}\(\varepsilon^\prime\)^2\right.\nn\\
& &\hspace{15mm}
\left. +\varepsilon^{1/2}\varepsilon^{\prime\prime}\(E^x\)^{-1} -
  \(E^x\)^{-2}\varepsilon^{1/2}\varepsilon^{\prime}\(E^x\)^{\prime}\]\,,
\end{eqnarray}
while keeping the diffeomorphism constraint unmodified.  In the classical
case, $f_1\(K_x\) = K_x^2$ and $f_2(K_x) = K_x$.  Here, we assume pointwise
holonomy corrections along the homogeneous directions while working in an
effective formalism. However, as shown in \cite{EffConsQBR}, adding additional
quantum moment terms does not change the structure of the constraint brackets
(while the constraints themselves usually do have moment corrections). By
keeping these modification functions general, we are able to examine the
restrictions imposed on them such that the modified constraints still have
closed brackets.

It is straightforward to see that the brackets between two diffeomorphism
constraints and between a Hamiltonian and a diffeomorphism constraint have the
same form as in the classical case.  The only complicated Poisson bracket is
thus the one between two Hamiltonian constraints, which  gives
\begin{eqnarray}
\{H[N_1], H[N_1]\} &=& \frac{1}{2G}\int \d\theta \(\,
2\frac{\varepsilon}{\(E^x\)^2}\(\frac{\d f_2}{\d K_x}\)\[K_x\(E^x\)^\prime -
\varepsilon^\prime\mathcal{A}\]\right.\nn\\
& &\hspace{5mm} \left. \frac{\varepsilon^\prime}{E^x}\[\(\frac{\d f_1}{\d
    K_x}\) - 2f_2\]\)\,,
\end{eqnarray}
where we have integrated by parts several times. On analysing this result, we
note two features:
\begin{enumerate}
\item The closure of the algebra is ensured only if we have $\d f_1/\d K_x -
  2f_2=0$, implying restrictions on the modification functions which have been
  kept free in the discussion so far. The coefficient of this term is neither
  the Hamiltonian nor the diffeomorphism constraint and thus would
  give rise to an anomaly term unless its coefficient vanishes.
\item Although closure can be ensured in this model by making the above
  restriction on the form of the holonomy modification functions, we obtain a
  structure function in the quantum theory which is deformed by a factor of
  $\d f_2/\d K_x$ as compared with the classical case. Using the
  consistency condition between $f_1$ and $f_2$, the factor takes the form $\d
  f_2/\d K_x= \frac{1}{2}\d^2f_1/\d^2K_x$.
\end{enumerate}
We thus have a deformation in the constraint algebra
\begin{eqnarray}\label{DeformAlg}
\left\{H[N_1], H[N_1]\right\} &=& \frac{1}{G}\int \d\theta \,
\frac{\varepsilon}{\(E^x\)^2}\frac{1}{2}\(\frac{\d^2 f_1}{\d
  K_x^2}\)\[K_x\(E^x\)^\prime - \varepsilon^\prime\mathcal{A}\]\,.
\end{eqnarray}
Signature change can be understood from this relation as follows: In models of
loop quantum gravity, holonomy modifications replace quadratic appearances of
extrinsic-curvature components in the Hamiltonian constraint by some bounded
functions which reach their maximum value near the Planck scale. The bounded
nature of these modification functions is a crucial ingredient in claims of
singuarity resolution in these models. Near a local maximum of a function such
as $f_1$, the second derivative is negative, making the right-hand side of
(\ref{DeformAlg}) change its sign. The same change of sign happens if one
switches the signature of the theory to Euclidean, and indeed the form of the
brackets has a close relationship with the hyperbolic or elliptic nature of
equations of motion consistent with the brackets \cite{Action,SigChange}.  A
negative correction factor in structure functions of (\ref{DeformAlg}) can
therefore be interpreted as indicating signature change. For $f_1\(K_x\) =
K_x^2$, on the other hand, we recover the classical result where the
modification in the structure function goes to one. Thus, in addition to
having a closed algebra for the modified constraints, we also recover the
hypersurface-deformation brackets in the classical limit. The model is
covariant provided our conditions are fulfilled. Only one free function,
$f_1(K_x)$, then remains, which is unrestricted by anomaly freedom and
covariance.

In \cite{GowdyAbel}, a loop quantisation of the LRS Gowdy model has been
proposed. To this end, the authors first Abelianise the classical bracket of
two normal deformations while leaving the other two relations
unchanged. Following \cite{LoopSchwarz}, the new, Abelianized constraint is
defined as a linear combination of the old Hamiltonian constraint and the
diffeomorphism constraints, while the diffeomorphism constraint remains
unchanged. (This partial Abelinisation can also be applied to the full
polarised Gowdy model \cite{GowdyCov} without local rotational symmetry.) The
new constraint used in this context, Eq.~(1) of \cite{GowdyAbel}, is
\begin{eqnarray}\label{newHam}
H_{\rm new}[N]=\frac{1}{2G}\int \d\theta\,
\frac{N}{\varepsilon^\prime} \[2\sqrt{\varepsilon} K_x^2 -
\frac{\sqrt{\varepsilon}\varepsilon^\prime}{2\(E^x\)^2} \]^\prime\,,
\end{eqnarray}
while $D[N^x]$ follows from (\ref{Diffeo}), as before.

The authors then adopt the holonomy modification scheme for models of loop
quantum gravity and substitute $K_x \rightarrow \sin{(\gamma K_x)}/\gamma$ in
(\ref{newHam}). The $K_x^2$ term in (\ref{newHam}) is therefore replaced by
$\(\sin{(\gamma K_x)}\)^2/\gamma^2$. The new constraint commutes with itself,
which is easy to see if we integrate by parts in (\ref{newHam}) (after
absorbing the denominator $\varepsilon^\prime$ in the lapse function ) and
notice that there are no spatial derivatives of $E^x$ anymore.  Although the
resulting theory is consistent in the sense of being anomaly-free, it is not
guaranteed to be covariant.  In order to show covariance, one must be able to
recover suitable generators of gauge transformations such that their brackets
lead to the hypersurface-deformation brackets in the classical limit. This
important conceptual step is missing in \cite{GowdyAbel}, but is completed
here.

We can start from (\ref{Ham}), having incorporated the holonomy modification
functions, and try to partially Abelianize this bracket. Thus, we first
holonomy-modify and then Abelianise. It is important to emphasise that we do
not impose any restrictions on either of the functions $f_1$ or $f_2$ at this
point. Proceeding as in the classical case, the new constraint is defined as
\begin{eqnarray}\label{NewHamDef}
\mathcal{H}_{\rm new}= \mathcal{H}_{\rm old} - \(\frac{\d f_1}{\d
  K_x}\)\sqrt{\varepsilon}\varepsilon^\prime \mathcal{D}\,,
\end{eqnarray}
where $\mathcal{H}, \mathcal{D}$ stand for the unsmeared versions of the
constraints. With this step, we arrive at the same form of the new,
holonomy-modified constraint as proposed in \cite{GowdyAbel}, {\it provided
  the two modification functions obey the condition $\d f_1/\d K_x
  = 2f_2$}. We have the same restriction on the modification functions as
found before by an analysis of anomaly freedom of hypersurface-deformation
brackets. Thus, requiring the new system of constraints to be (partially)
Abelian is equivalent to imposing that the old system of constraints form a
closed system. The closed hypersurface deformation brackets then again
indicate signature change.

We can arrive at this result from another perspective as well. Starting with
the newly defined classical constraint (\ref{newHam}), one can introduce a
quantum theory as in \cite{GowdyAbel}. However, to ensure covariance we must
be able to define constraints which have hypersurface-deformation brackets
with the correct classical limit. This condition translates to recovering a
Hamiltonian constraint from the Abelianized constraint by inverting the linear
transformation used above, which can be equivalently thought of as
transforming the lapse function and the shift vector as
\begin{eqnarray}\label{LapseShift}
N &=& \frac{\tilde{N}}{E^x}\\
N^\theta &=& \tilde{N}^\theta - \tilde{N}\frac{\sqrt{\varepsilon}\:\d f_1/\d
    K_x}{\varepsilon^\prime}\,.
\end{eqnarray}
This step puts the system of constraints in the form of our ansatz (\ref{Ham})
and (\ref{Diffeo}), with the specific choice of $f_1=\(\sin{(\gamma
  K_x)}\)^2/\gamma^2$. As expected, for the holonomy-modified LRS Gowdy system
in \cite{GowdyAbel}, signature change occurs in high curvature regions: The
second derivative of $f_1$ in this case is proportional $-2\cos(2\gamma K_x)$,
which has a negative sign near a local maxima of $f_1$.

It is remarkable that our result and signature change are robust even when
different equivalent systems of classical constraints are used as the starting
point of a loop quantization. As demonstrated earlier with spherical symmetry
\cite{SphSymmCov}, the restrictions on holonomy modification functions are the
same, no matter whether they are derived by requiring closure of the algebra
or by requiring that it be possible to define new constraints which have
partially Abelian brackets. The present section shows that this conclusion is
also true for another model of loop quantum gravity, namely the LRS Gowdy
model. We have shown that signature change is an unavoidable consequence of
holonomy modifications in this model, irrespective of how one defines the
system of constraints as long as one forces the resulting quantum theory to
be covariant. This result may be taken as an indication that these conclusions
hold more generally in midisuperspace models of loop quantum gravity without
local physical degrees of freedom. The remainder of this paper confirms
this expectation.

\section{General case}

A theory without local degrees of freedom should have as many pairs of
canonical variables as there are first-class constraints. For
hypersurface-deformation covariant systems in two space-time dimensions, there
should therefore be two pairs of canonical fields, which we continue to denote
as in the LRS Gowdy model of the preceding section. A generic form of a
Hamiltonian constraint is
\begin{eqnarray}
H[N]&=&-\frac{1}{2G}\int \d\theta N\(\theta\) \left\{f\(\A, K_x, E^x,
  \varepsilon\) +g_1(\varepsilon) \frac{(\varepsilon^\prime)^2}{E^x}
  +g_2(\varepsilon) \frac{\varepsilon^{\prime\prime}}{E^x}\right.\nn\\
 & &\hspace{3cm}\left.+g_3(\varepsilon) \frac{\varepsilon^\prime
     (E^x)^\prime}{(E^x)^2} +g_4(\varepsilon) E^x\right\}\,,\label{GenHam}
\end{eqnarray}
whereas the diffeomorphism constraint again has the usual form
\begin{eqnarray}
D[N^\theta] = \frac{1}{G} \int \d\theta N^\theta(\theta)\{K_x^\prime E^x -
\varepsilon^\prime \A \}
\end{eqnarray}
if the spatial structure remains unchanged.  The Poisson brackets between the
canonical variables remain the standard ones $\{K_x(x), E^x(y)\} = G
\delta(x,y) = \{\A(x), \varepsilon(y)\}$. One might expect quantum corrections
in the Poisson structure, but by Darboux' theorem one can always transform
back to canonical variables. All such corrections are then contained in the
modification functions already introduced. (The structure of the
diffeomorphism constraint is strongly restricted for canonical variables and
would not change by such a transformation.)

The assumptions for our general form are:
\begin{enumerate}
\item The diffeomorphism constraint does not have modifications. For models of
  loop quantum gravity, this assumption is made because one usually quantizes
  the diffeomorphism constraint, or rather the finite action it generates,
  without taking recourse to holonomies around loops. (For an exception see
  \cite{DiffeoOp}.)
\item All curvature dependence is contained in a generic function $f$, while
  spatial derivatives of the triad components have separate correction
  functions. One could include the last term $g_4(\epsilon)E^x$ in the
  function $f$, but it is more convenient to keep it separate.
\item Every term in (\ref{GenHam}) has the correct density weight as
  required. (See \cite{SphSymm} for a discussion of density weights in
  midisuperspace models.) We do not consider terms with spatial derivatives
in the denominator because they would not be guaranteed to be finite
everywhere.
\item There are no terms of higher than second spatial derivatives to
  the order considered here. Such terms would require a derivative expansion
  as in \cite{HigherSpatial}.
\item In midisuperspace models of general relativity, terms proportional to
  the second order derivatives of $E^x$ are absent due to the fact that
  spatial derivatives come from the curvature tensor which cannot have two
  radial derivatives of the radial components owing to its antisymmetry
  properties. Thus we do not have terms proportional to $(E^x)^{\prime\prime}$
  or $((E^x)^{\prime})^2$. In Sec.~\ref{s:New}, we will show that such terms
  are, in fact, impossible in an anomaly-free system of the form
  (\ref{GenHam}).
\end{enumerate}

Our goal is to start with this ansatz and try to impose conditions on the
arbitrary functions by requiring closure of the constraint algebra. We will
also impose that the Hamiltonian constraint has the correct classical limit
for small curvature components and large $\varepsilon$. Both conditions taken
together then ensure covariance.

\subsection{Brackets}

Looking at the $\{H,H\}$ bracket, we know that the only non-zero contributions
come from the first term with the rest of the terms in the Hamiltonian
constraint.  We write each of these contributions from $\{H[N_1], H[N_2]\}$
individually. From now on, we are going to suppress the functional dependence
of each of these arbitrary functions on the canonical variables. Term $1$
with term $5$ gives a vanishing contribution. Term $1$ with term $2$
gives
\begin{eqnarray}
-\frac{1}{2 G}\int \d\theta \(N_1 N_2^\prime - N_1^\prime N_2\)
\(\(\frac{\partial f}{\partial \A}\) g_1
\(\frac{\varepsilon^\prime}{E^x}\)\)\,.\label{1with2}
\end{eqnarray}
Term $1$ with term $4$ gives
\begin{eqnarray}
-\frac{1}{4 G}\int \d\theta \(N_1 N_2^\prime - N_1^\prime N_2\)
\(\(\frac{\partial f}{\partial K_x}\)g_3
\(\frac{\varepsilon^\prime}{(E^x)^2}\) + \(\frac{\partial f}{\partial
  \A}\)g_3\(\frac{(E^x)^\prime}{(E^x)^2}\)\)\,.\label{1with4}
\end{eqnarray}
Finally, term $1$ with term $3$ gives
\begin{eqnarray}
& & \frac{1}{4 G}\int \d\theta \(N_1 N_2^\prime - N_1^\prime N_2\)
\left(\(\frac{\partial f}{\partial
    \A}\)\dot{g}_2\(\frac{\varepsilon^\prime}{E^x}\) - \(\frac{\partial
    f}{\partial \A}\)g_2\(\frac{(E^x)^\prime}{(E^x)^2}\) \right.\nn\\
& & \left. - \(\frac{\partial^2 f}{\partial \A \partial
    K_x}\)g_2\(\frac{K_x^\prime}{E^x}\) - \(\frac{\partial^2 f}{\partial
    \A^2}\)g_2\(\frac{\A^\prime}{E^x}\) - \(\frac{\partial^2 f}{\partial
    \A \partial E^x}\)g_2\(\frac{(E^x)^\prime}{E^x}\) - \(\frac{\partial^2
    f}{\partial \A \partial
    \varepsilon}\)g_2\(\frac{\varepsilon^\prime}{E^x}\)\right)\,.\label{1with3}
\end{eqnarray}
In the above expressions, a dot above any function dependent on a single
variable refers to its derivative with respect to its variable.

The requirement for the algebra to be closed implies that any bracket between
two constraints must be another constraint. This means that the right-hand
side of $\{H,H\}$ can, in addition to the diffeomorphism constraint, also
include a Hamiltonian constraint, provided its coefficient goes to zero in the
classical limit. It turns out that just looking at conditions for the
diffeomorphism constraint to appear, perhaps with modified structure
functions, results in strong conditions on the free functions.

Since there is no $\A^\prime$ term which can appear on the right-hand side, we
have $\partial^2 f/\partial \A^2 = 0$. Thus $f$ is linear in $\A$ and can be
written as
\begin{eqnarray}
f(\A,K_x,E^x,\varepsilon) = f_2(K_x,E^x,\varepsilon)\A +
f_3(K_x,E^x,\varepsilon)\,. \label{linearinA}
\end{eqnarray}
Similarly, there is no $(E^x)^{\prime}$ term on the right-hand side, implying
\begin{equation}
f_2 g_3 (E^x)^{-2} + f_2 g_2 (E^x)^{-2} + \(\frac{\partial f_2}{\partial
  E^x}\) g_2 (E^x)^{-1} = 0
\end{equation}
or, equivalently,
\begin{equation}
g_2 + g_3 + \frac{g_2}{f_2}\(\frac{\partial f_2}{\partial E^x}
E^x\) = 0\,.\label{A1}
\end{equation}
This expression can be rearranged to bring it to the form
\begin{eqnarray}
-\frac{g_2 + g_3}{g_2} = \frac{E^x}{f_2}\(\frac{\partial f_2}{\partial
  E^x}\)\,,\label{A}
\end{eqnarray}
where now the left-hand side depends only on $\varepsilon$ whereas the
right-hand side depends on $\varepsilon, E^x$ and $K_x$. Therefore both sides
must each be equal to the same function of $\varepsilon$, which we call
$g_5(\varepsilon)$. Therefore,
\begin{equation}
 1+g_3/g_2= -g_5
\end{equation}
and
\begin{equation}
\frac{E^x}{f_2}\(\frac{\partial f_2}{\partial E^x}\) = g_5\,.
\end{equation}
We conclude that
\begin{equation}
f_2\(K_x, E^x, \varepsilon\) = \tilde{f}_2\(K_x, \varepsilon\)
(E^x)^{-(1+g_3/g_2)}\,,\label{Aprime}
\end{equation}
where we restore the explicit definition of $g_5$ in the final line.

Going back to the expressions (\ref{1with2}), (\ref{1with3}) and
(\ref{1with4}), we notice that any term proportional to just
$\varepsilon^\prime$ (without a multiplicative factor of $\A$) must also be
set to zero since there is no such term in the diffeomorphism constraint:
\begin{equation}
2f_2g_1(E^x)^{-1} + g_3\(\frac{\partial f_3}{\partial K_x}\)(E^x)^{-2} -
f_2\dot{g}_2(E^x)^{-1} + g_2 \(\frac{\partial f_2}{\partial \varepsilon}\)
(E^x)^{-1} = 0
\end{equation}
or
\begin{equation}
 \[2f_2g_1 -f_2\dot{g}_2 +g_2\(\frac{\partial f_2}{\partial
  \varepsilon}\)\] E^x = -g_3 \(\frac{\partial f_3}{\partial
  K_x}\)\,.\label{B}
\end{equation}
On the left-hand side, we can use (\ref{Aprime}) to write out the dependence
of the expression on $E^x$. Since the right-hand side involves $g_3$, which is
a function of $\varepsilon$ alone, and the derivative of $f_3$ with respect to
$K_x$, we can deduce that the dependence of $f_3$ on $E^x$ is
\begin{eqnarray}
f_3(K_x, E^x, \varepsilon) = \tilde{f}_3(K_x, \varepsilon)
(E^x)^{-g_3/g_2}\,. \label{B1}
\end{eqnarray}
Inserting (\ref{Aprime}) and (\ref{B1}) in (\ref{B}),
\begin{eqnarray}
2\tilde{f}_2g_1 -\tilde{f}_2\dot{g}_2 + g_2\(\frac{\partial
  \tilde{f}_2}{\partial \varepsilon}\) = -g_3 \(\frac{\partial
  \tilde{f}_3}{\partial K_x}\)\,.\label{Bprime}
\end{eqnarray}

Looking at the remaining two terms left, one of which is proportional to
$\A\varepsilon^\prime$ and the other to $K_x^\prime E^x$, we have
\begin{eqnarray}
\frac{\partial f_2}{\partial K_x}\[g_2 (E^x)^{-2}\(K_x^\prime E^x\) + g_3
(E^x)^{-2} \(\varepsilon^\prime\A\)\]\,.\label{C+D}
\end{eqnarray}
For this to be proportional to the diffeomorphism constraint, we require that
the prefactor of both the $K_x^\prime E^x$ and the $\A\varepsilon^\prime$ be
the same. This implies
\begin{eqnarray}
g_2 = -g_3\,.\label{I}
\end{eqnarray}
We can use this relation in (\ref{Aprime}) and (\ref{B1}),
\begin{eqnarray}\label{II}
f_2(K_x, E^x, \varepsilon) &=& \tilde{f}_2(K_x, \varepsilon)\,,\\
f_3(K_x, E^x, \varepsilon) &=& \tilde{f}_3(K_x, \varepsilon) E^x\,.
\end{eqnarray}
From (\ref{Bprime}) and (\ref{I}),
\begin{eqnarray}\label{IIIprime}
2\tilde{f}_2g_1 - \tilde{f}_2\dot{g}_2 + g_2\(\frac{\partial
  \tilde{f}_2}{\partial \varepsilon}\) = g_2\(\frac{\partial
  \tilde{f}_3}{\partial K_x}\)
\end{eqnarray}

\subsection{Implications and special cases}

Some of our new relations have interesting interpretations, which we collect
in this subsection.

Equation (\ref{I}) implies that the two terms
\begin{equation}
 g_2(\varepsilon) \frac{\varepsilon^{\prime\prime}}{E^x}+g_3(\varepsilon)
 \frac{\varepsilon^\prime  (E^x)^\prime}{(E^x)^2} = g_2(\varepsilon)
 \left(\frac{\varepsilon^{\prime\prime}}{E^x}- \frac{\varepsilon^\prime
     (E^x)^\prime}{(E^x)^2}\right)= -2 g_2(\varepsilon) \Gamma'
\end{equation}
can always be written in terms of the classical spin connection component
\begin{equation}
 \Gamma = -\frac{\epsilon'}{2E^x}
\end{equation}
of a midisuperspace metric.

Equation (\ref{C+D}) shows that the structure function of the modified system
is equal to
\begin{equation}
 \frac{\partial f_2}{\partial K_x} \frac{g_2}{(E^x)^2}= \beta
 \frac{\varepsilon}{(E^x)^2}
\end{equation}
with the modification function
\begin{equation}
 \beta= \frac{\partial f_2}{\partial K_x} \frac{g_2}{\varepsilon}\,.
\end{equation}
Using (\ref{IIIprime}), $\partial f_2/\partial K_x$ is proportional to
$\partial^2f_3/\partial K_x^2$ if the dependence of $f_2$ on $\varepsilon$ is
weak. Around a local maximum of $f_3$ in $K_x$, the modification function
$\beta$ is therefore negative and we obtain signature change.

The modification function $\beta$ does not introduce a dependence of structure
functions on $g_4$, and there is no restriction on $g_4$ from anomaly
freedom. There should therefore be classical gravity models for any choice of
$g_4(\varepsilon)$. Indeed, as the canonical transformation derived in
\cite{SphSymmPSM} shows, if $g_4$ is the only modification function that
differs from spherical symmetry, (\ref{GenHam}) is nothing but a 2-dimensional
dilaton model with potential $V(\varepsilon)=g_4(\varepsilon)$, expressed in
connection variables as used in models of loop quantum gravity. (The function
$g_1$ does not appear explicitly in the expression of $\beta$, but it cannot
be chosen independently because it is related to $f_2$, $f_3$ and $g_2$ by
(\ref{IIIprime}).)

It is not easy to analyze Eq.~(\ref{IIIprime}) in general form, but a few
special cases are of interest. First, we can see that it is not compatible
with lattice refinement \cite{InhomLattice,CosConst} which would require a
dependence of modification functions on extrinsic curvature via the
combination $\epsilon^qK_x$ with some real number $q$. If we assume two
different such dependences in $\tilde{f}_2(\epsilon^pK_x)$ and
$\tilde{f}_3(\epsilon^qK_x)$, (\ref{IIIprime}) implies
\begin{equation}
 2\tilde{f}_2 g_1- \tilde{f}_2 \dot{g}_2+ p\epsilon^{p-1} g_2 K_x
 \dot{\tilde{f}}_2 = g_2\epsilon^q \dot{\tilde{f}}_3\,.
\end{equation}
The third term with a factor of $K_x$ is incompatible with almost periodic
functions $\tilde{f}_2$ and $\tilde{f}_3$ as assumed in models of loop quantum
gravity.

Another special case is given by a factorizable ansatz for the modification
functions:
\begin{eqnarray}\label{separationofvariables}
\tilde{f}_2\(K_x, \varepsilon\) &=& f_4(K_x) g_6(\varepsilon)\\
\tilde{f}_3\(K_x, \varepsilon\) &=& f_5(K_x) g_7(\varepsilon)\,.
\end{eqnarray}
Inserting this form in (\ref{IIIprime}), we find
\begin{eqnarray}
2\frac{g_6g_1}{g_7g_2} - \frac{g_6\dot{g}_2}{g_7g_2} + \frac{\dot{g}_6}{g_7} =
\frac{\dot{f}_5}{f_4}\,.
\end{eqnarray}
The left-hand side depends only on $\varepsilon$ while the right-hand side
depends solely on $K_x$. Thus, each of the two sides must be equal to a
constant.
\begin{eqnarray}
\frac{\d f_5}{\d K_x} = c f_4\,,\label{III}
\end{eqnarray}
and
\begin{eqnarray}
2\frac{g_6g_1}{g_7g_2} - \frac{g_6\dot{g}_2}{g_7g_2} + \frac{\dot{g}_6}{g_7} =
c\,.\label{IV}
\end{eqnarray}

The form of our generalized Hamiltonian constraint is now restricted to be
\begin{eqnarray}
H[N]= -\frac{1}{2G}\int\d\theta N(\theta) \left\{g_7 f_5(K_x) E^x + g_6
  f_4(K_x)\A\right.\nn\\
\left. \,\,\,\,\, +g_1\(\frac{(\varepsilon^\prime)^2}{E^x}\) +
  g_2\(\frac{\varepsilon^{\prime\prime}}{E^x}\) - g_2
  \(\frac{\varepsilon^\prime(E^x)^\prime}{(E^x)^2}\)
  +g_4E^x\right\}\,.\label{Hamalmostfinal}
\end{eqnarray}
All of the $g$-functions are functions of $\varepsilon$ with their functional
dependence suppressed. However, not all of the remaining functions are
unconstrained. We have the additional conditions given in (\ref{III}) and
(\ref{IV}). We can also absorb $g_7$ in the lapse function and rescale the
rest of the $g$-functions accordingly. In other words, we can set $g_7=1$
without any loss of generality. We call the new lapse function $\tilde{N}$.

For our generalised midisuperspace model, closure of two Hamiltonian
constraints, including holonomy modifications, implies the condition
(\ref{III}) for the modification functions. Given this condition, the
deformed structure function takes the form
\begin{eqnarray}\label{ModStrucFunc}
\frac{1}{c}\(\frac{\d^2 f_5}{\d K_x^2}\) g_2 g_6 (E^x)^{-2}\,,
\end{eqnarray}
while the final form of the Hamiltonian constraint is
\begin{eqnarray}
H[N]= -\frac{1}{2G}\int\d\theta \tilde{N}(\theta) \left\{ f_5(K_x) E^x +
  \(\frac{g_6}{c}\)\(\frac{\d f_5}{\d K_x}\) \A\right.\nn\\
\left. \,\,\,\,\, +g_1\(\frac{(\varepsilon^\prime)^2}{E^x}\) +
  g_2\(\frac{\varepsilon^{\prime\prime}}{E^x}\) - g_2
  \(\frac{\varepsilon^\prime(E^x)^\prime}{(E^x)^2}\)
  +g_4E^x\right\}\,.\label{Hamfinal}
\end{eqnarray}
The classical limit is given by
$g_2(\varepsilon)g_6(\varepsilon)=\varepsilon$, while $f_5(K_x)=K_x^2$, with
$c=2$.  The function $g_4$ then labels different classical models with
undeformed covariance, including all 2-dimensional dilaton gravity models, or
the spherically symmetric model as well as Gowdy LRS.

\subsection{Second-order spatial derivatives beyond general relativity}
\label{s:New}

In our analysis so far, we did not consider two terms proportional to
second-order spatial derivatives of triad components, namely $(E^x)''$ and
$((E^x)')^2$. These terms do not arise in midisuperspace models of general
relativity due to antisymmetry properties of the Riemann curvature tensor
since derivatives with respect to the radial coordinate hence cannot appear on
the radial component of the triads. However, such terms could conceivably
arise if there is some modification to general relativity. Here, we show that
the presence of such terms is incompatible with having anomaly-free
constraints.

Taking into account density weights, the general form of the Hamiltonian
constraint with the additional terms is given by
\begin{eqnarray}
H[N]&=&-\frac{1}{2G}\int \d\theta N\(\theta\) \left\{f\(\A, K_x, E^x,
  \varepsilon\) +g_1(\varepsilon) \frac{(\varepsilon^\prime)^2}{E^x}
  +g_2(\varepsilon) \frac{\varepsilon^{\prime\prime}}{E^x}\right.\\
 & &\hspace{3cm}\left.+g_3(\varepsilon) \frac{\varepsilon^\prime
     (E^x)^\prime}{(E^x)^2} +g_4(\varepsilon) E^x
   +h_1(\varepsilon)\frac{(E^x)^{\prime\prime}}{(E^x)^2}
   +h_2(\varepsilon)\frac{(E^{x\prime})^2}{(E^x)^3}
 \right\}\nn\label{MoreGenHam}
\end{eqnarray}
with two new functions $h_1(\varepsilon)$ and $h_2(\varepsilon)$.
The new terms arising from the Poisson bracket of two such Hamiltonian
constraints are
\begin{equation}\label{newterms}
-\frac{1}{2G}\int \d \theta (N_1N_2^\prime - N_1^\prime
N_2) \[\(\frac{\partial f}{\partial K_x}\) h_2 E^{x\prime}  (E^x)^{-3}\]
\end{equation}
and
\begin{eqnarray} \label{newterms2}
& & -\frac{1}{4G}\int \d \theta (N_1N_2^\prime - N_1^\prime N_2)
\left[2\(\frac{\partial f}{\partial K_x}\)h_1  E^{x\prime}  (E^x)^{-3} -
  \(\frac{\partial f}{\partial K_x}\) \dot{h_1} \varepsilon^\prime (E^x)^{-2}
\right.\nn\\
& & \hspace{2cm}\left. +\(\frac{\partial^2 f}{\partial K_x^2}\)h_1 K_x^\prime
  (E^x)^{-2} + \(\frac{\partial^2 f}{\partial K_x \partial \A}\) h_1 \A^\prime
  (E^x)^{-2}\right.\nn\\
& & \hspace{2cm} \left. + \(\frac{\partial^2 f}{\partial K_x \partial
    E^x}\)h_1E^{x\prime} (E^x)^{-2} + \(\frac{\partial^2 f}{\partial
    K_x \partial \varepsilon}\)h_1 \varepsilon^\prime (E^x)^{-2}\right]\,.
\end{eqnarray}
These new terms contribute to all the conditions we had before. Starting
with the requirement that there be no terms proportional to $\A^\prime$ on the
right-hand side, we have
\begin{eqnarray}\label{noAprime}
g_2E^x\(\frac{\partial^2 f}{\partial \A^2}\) +h_1 \(\frac{\partial^2
  f}{\partial \A \partial K_x}\) = 0\,.
\end{eqnarray}
Defining $f_1(\A, K_x, \varepsilon, E^x):= \partial f/\partial \A$,
\begin{eqnarray}
g_2E^x\(\frac{\partial f_1}{\partial \A}\) =-h_1 \(\frac{\partial
  f_1}{\partial K_x}\)\,.
\end{eqnarray}
We can solve this equation by
\begin{equation}
 f_1({\cal A},K_x,E^x,\varepsilon) = F(K_x-{\cal A} h_1/(g_2E^x),
 E^x,\varepsilon)
\end{equation}
with an arbitrary function $F$ of three variables.  Since $f_1=\partial
f/\partial {\cal A}$, we have
\begin{equation} \label{fGH}
 f({\cal A},K_x,E^x,\varepsilon) = G(K_x-{\cal A} h_1/(g_2E^x),
 E^x,\varepsilon)+ H(K_x,E_x,\varepsilon)
\end{equation}
with $\partial G/\partial {\cal A}=F$ and another free function
$H$ of three arguments.

We can already see that the new terms are likely to lead to problematic
conditions on the modification functions: The component ${\cal A}$ can only
appear in the specific combination $K_x-{\cal A} h_1/(g_2E^x)$ with $K_x$, but
finding anomaly-free modifications of the ${\cal A}$-dependence has proven
difficult \cite{HigherSpatial}. The function $G$ could then only be a linear
function in its first argument.

In fact, the new terms are ruled out if we use (\ref{fGH}) and evaluate
all contributions to the bracket that could give rise to the term ${\cal
  A}\varepsilon'$ in the diffeomorphism constraint. In particular, we have to
make sure that we have a factor of ${\cal A}$ but no factor of $K_x$
multiplying $\varepsilon'$. Two such terms,
\begin{equation}
 \frac{\partial^2f}{\partial{\cal A}\partial\varepsilon} \frac{g_2}{E^x} +
 \frac{\partial^2f}{\partial K_x\partial\varepsilon} \frac{h_1}{(E^x)^2}=
 \frac{\partial^2H}{\partial K_x\partial\varepsilon} \frac{h_1}{(E^x)^2}
\end{equation}
do not contribute a factor of ${\cal A}$ as coefficients of
$\varepsilon'$. The remaining terms are
\begin{equation}
 -\frac{\partial g}{\partial K_x} \frac{g_3}{(E^x)^2}+ \frac{\partial
   f}{\partial {\cal A}} \frac{\dot{g}_2}{E^x}+ \frac{\partial f}{\partial
   K_x} \frac{\dot{h}_1}{(E^x)^2} = -G_1 (g_3+h_1\dot{g}_2/g_2-\dot{h}_1)
\end{equation}
plus terms that do not depend on ${\cal A}$, where $G_1$ is the partial
derivative of $G$ by its first argument. We obtain a coefficient with linear
dependence on ${\cal A}$ only if $G$ is quadratic in $K_x-{\cal A}
h_1/(g_2E^x)$, but even if this is the case, there will be additional terms
depending on $K_x$ which do not all cancel out. It is therefore impossible to
gather all the new terms in coefficients of the diffeomorphism constraint, and
no anomaly-free formulation is possible unless $h_1=0$.

With this result, we can follow the previous steps up to Eq.~(\ref{A1}). There
is now a new term $h_2 (E^x)^{-3} \partial f/\partial K_x$ in the resulting
equation
\begin{eqnarray}\label{noExprime}
& &\(\frac{\partial f}{\partial \A}\)g_3(E^x)^{-2} + \(\frac{\partial
  f}{\partial \A}\)g_2(E^x)^{-2} + \(\frac{\partial^2 f}{\partial \A \partial
  E^x}\) g_2(E^x)^{-1}
\nn\\
& &\,\,+ 2\(\frac{\partial f}{\partial K_x}\)h_2(E^x)^{-3}  = 0
\end{eqnarray}
which, for $f$ of the form (\ref{linearinA}), contains a factor of ${\cal
  A}$. However, all other terms in (\ref{noExprime}) are independent of ${\cal
  A}$, which is compatible with the new term only if $\partial f_2/\partial
K_x=0$. But in this case there is no term of the form $K_x{\cal A}$ in
the Hamiltonian, and the model is not compatible with the classical
limit. Therefore, $h_2=0$ and both new terms are ruled out.

\section{Conclusions}

We have analyzed a general canonical form of 2-dimensional covariant
models without local physical degrees of freedom. A large subclass of
such models has been recognized as \textit{classical} 2-dimensional
dilaton gravity models with an arbitrary potential. Another large
class of models, most of which have not been encountered before, has a
deformed notion of covariance and includes models of loop quantum
gravity. Holonomy-modified versions of the 2-dimensional dilaton
gravity models, as studied for instance in \cite{AbelDil,AbelCGHS},
fall within the latter group.  In this class, signature change is a
generic consequence of modifications that introduce a bounded
dependence of the Hamiltonian constraint on extrinsic curvature.

Our results unify several recent investigations of midisuperspace models of
loop quantum gravity, including \cite{GowdyAbel,AbelDil,AbelCGHS}. They also
provide further support for the genericness of signature change in models of
loop quantum gravity. So far, signature change has been avoided only by
following three distinct procedures: (i) Using classical assumptions on the
structure of space-time and foregoing an analysis of anomaly freedom. (ii)
Implementing modifications via canonical transformations
\cite{CollapseQG}. (iii) Using complex connections
\cite{SphSymmComplex,CosmoComplex}. The first option is problematic because it
does not guarantee anomaly freedom. The second option is problematic as well,
as discussed in the appendix. The third option needs to be explored further,
in particular regarding the implementation of reality conditions. Furthermore, for complex variables, the quantization scheme becomes rather important since holonomy corrections within certain programs can still lead to signature change \cite{MDRfromLQG}.

\section*{Acknowledgements}

This work was supported in part by NSF grant PHY-1607414.

\begin{appendix}

\section{An alternative modification scheme}

A new ``modification'' scheme with bounded functions of curvature has been
proposed in \cite{CollapseQG}, using spherically symmetric models. Instead of
modifying the Hamiltonian constraint, the authors use a canonical
transformation $K_x \rightarrow f(K_x)$ and $E^x \rightarrow
E^x/\dot{f}(K_x)$, where the dot again represents the derivative of the
function with respect to its argument, here $K_x$. (We have translated the
relations of \cite{CollapseQG} to the notation used in the main body of the
present paper. Instead of $K_{\varphi}$ in \cite{CollapseQG}, we therefore
write $K_x$.) In the specific case of \cite{CollapseQG}, the function is
chosen as the usual sine function of models of loop quantum gravity, but we
choose to keep the analysis more general. The Poisson brackets indeed remain
unchanged:
\begin{eqnarray}
\left\{f(K_x(x))\,,\,\frac{E^x}{\dot{f}(K_x)}(y) \right\} =
\left\{K_x(x)\,,\,E^x(y) \right\} = \frac{1}{2}G\delta(x,y) \,.
\end{eqnarray}

\subsection{Constraints}

Starting with the classical Hamiltonian and diffeomorphism constraints, the
canonical transformation takes us to
\begin{eqnarray}
H[N]&=&-\frac{1}{2G}\int \mathrm{d}x \,\,N\Bigg[({{\varepsilon}})^{-1/2}
\(\frac{E^{x}}{\dot{f}(K_x)}\)f(K_x)^2 +
2({{\varepsilon}})^{1/2}{\cal A}f(K_x)
+({{\varepsilon}})^{-1/2}\(\frac{E^{x}}{\dot{f}(K_x)}\) \nn \\
& & - \frac{1}{4}({{\varepsilon}})^{-1/2}(\varepsilon')^{2}\(\frac{E^{x}}
{\dot{f}(K_x)}\)^{-1} - ({{\varepsilon}})^{1/2}\varepsilon''
\(\frac{E^{x}}{\dot{f}(K_x)}\)^{-1} \nn\\
& & + ({{\varepsilon}})^{1/2}\varepsilon'\(\frac{E^{x}}{\dot{f}(K_x)}\)^{-2}
\left(\frac{E^{x\prime}}{\dot{f(K_x)}} -
  \frac{E^x\ddot{f(K_x)}K_x'}{(\dot{f(K_x)})^2}\right)\Bigg]. \\
D[N^x]&=&\frac{1}{2G}\int\mathrm{d}x \,\,N^x\left[2{E^x} K'_x-{\cal A}
  \varepsilon'\right]\,.
\end{eqnarray}
Viewed as a modified expression, this $H[N]$ has not been included in our main
analysis because it would require modification functions $g_i(K_{\varphi})$
that do not just depend on $\varepsilon$.

Following the procedure outlined in the previous sections, we can calculate
the Poisson bracket between these constraints and find that the constraint
algebra takes the form
\begin{eqnarray}
\left\{D[N^x], D[M^x]\right\} &=& D[\mathcal{L}_{N^x}M^x],\\
 \left\{H[N], D[N^x]\right\} &=& -H[\mathcal{L}_{N^x}N],\\
 \left\{H[N], H[M]\right\} &=&
 D[(NM'-MN')|{{\varepsilon}}|\(\frac{E^{x}}{\dot{f}(K_x)}\)^{-2}]\,.
\end{eqnarray}
(More details of the derivation are given in the following subsection.)
As expected, the new structure function agrees with the usual one after
applying the canonical transformation.  One could interpret the last bracket
as a hypersurface-deformation bracket with structure function modified by a
factor of $\dot{f}^2$. This function is positive and therefore does not lead
to signature change. According to the general results of \cite{Normal}, it can
therefore be absorbed by a field redefinition, which would just be the inverse
of the canonical transformation.

Once one (partially) Abelianizes the system of (modified) constraints,
following \cite{LoopSchwarz}, the Abelianized constraints remain Abelianized
in spite of the modifications \cite{CollapseQG}. In fact, even in the presence
of matter, the total constraints (gravitational plus the matter parts) form a
(partially) Abelianized algebra. However, if we go back to the original
hypersurface-deformation genrators, the structure functions are deformed, as
shown here.

This ``modification'' procedure suffers from several drawbacks. By applying a
canonical transformation to the classical constraints, one cannot arrive at
modified dynamics. (There is then no actual modification at all.) It is
surprising how \cite{CollapseQG} can nevertheless make claims about
singularity resolution. In fact, the canonical transformation is one-to-one
only in a range of $K_x$ where $f(K_x)$ is monotonic. For the common functions
used in models of loop quantum gravity, this excludes all values of $K_x$
greater than a certain finite threshold. The classical singularity (infinite
$K_x$) is eliminated from these models only because the canonical
transformation is valid only in a limited part of phase space.

Moreover, the form of the modification is in contradiction with the usual
guiding principles, which suggest modifications of curvature terms in the
Hamiltonian constraint, but no inverses of $\dot{f}(K_x)$ in triad terms.

\subsection{Derivation of the structure function}

The modification procedure introduced in \cite{CollapseQG} is of the form
\begin{eqnarray}
K_{x}  & \rightarrow & f(K_x)\\
E^x & \rightarrow & E^x/\dot{f}(K_x)\,.
\end{eqnarray}
Instead of showing the entire derivation of the constraint brackets after this
transformation, we give a brief sketch of the derivation below. There are
three types of terms which we shall be confronting during this
calculation. The crucial point is that, in this instance, the brackets of the
modified variables are the same for the classical ones, by construction.

The first type is of the form
\begin{eqnarray}
& &\left\{g_1g_2f(K_x) \(E^x/\dot{f}(K_x)\)\,,\, h_1h_2f(K_x)
  \(E^x/\dot{f}(K_x)\)\right\}\,.
\end{eqnarray}
(We use $g$ and $h$ to denote functions of the phase space variables arising
in the Hamiltonian constraint, while reserving $f$ for the modification
function.) These terms do not contribute to the $\{H , H\}$ bracket at all.

The second type of brackets which comes up in the calculation, are
\begin{eqnarray}
& &\left\{gf(K_x)\,,\,h\(E^x/\dot{f}(K_x)\)\,
  \(E^x/\dot{f}(K_x)\)^\prime\right\}\nn\\
&=& \frac{\d }{\d K_x}\(g\(K_x\)\) \frac{\d}{\d
  E^x}\(h\(E^{x}\)\)\,\frac{\partial}{\partial y}\(\delta(x,y)\) + \ldots
\end{eqnarray}
The dots above are terms which are proportional to the delta function
$\delta(x,y)$, but not to derivatives of the delta function. Such terms cancel
out of the $\{H , H\}$ bracket due to an opposite contribution.

Finally, we encounter terms of the form
\begin{eqnarray}
\left\{g\(E^x/\dot{f}(K_x)\) \,,\,h\[\(E^x/\dot{f}(K_x)\)^\prime\]\right\} \,.
\end{eqnarray}
Such terms for the unmodified variables were of the form $\{E^x ,
E^{x\prime}\}$, which was trivially zero. For the transformed variables,
\begin{eqnarray}
& &\left\{g\(E^x/\dot{f}(K_x)\) \,,\,h\[\(E^x/\dot{f}(K_x)\)^\prime\]\right\}=
0  + \ldots
\end{eqnarray}
The dots are again terms proportional to $\delta(x,y)$ which are
cancelled by opposite terms. However, the crucial point is that new terms
with derivatives of delta functions are generated in this case. However, they
cancel among each other.

Due to the above observations, and the fact that the Hamiltonian constraint is
linear in $E^{x\prime}$, we can simply replace the structure function written
in terms of the classical phase-space variables by its modified twin. In the
specific case of Gowdy LRS, just as in spherical symmetry, this implies
$\varepsilon/\(E^x\)^2 \rightarrow
\varepsilon\(\dot{f}(K_x)\)^2/\(E^x\)^2$. Similar arguments work even when a
matter contribution (say, in the form of a minimally coupled scalar field) is
taken into account. Once again, the structure functions appearing in the
brackets of the total constraints (gravitational plus the matter
contributions) have the same deformation as above.

\end{appendix}


\end{document}